\documentclass{ws-procs9x6}
\usepackage{epsfig}
\usepackage{amssymb}
\begin{document}
\bibliographystyle{ws-procs9x6}

\newcommand{\be}{\begin{eqnarray}}
\newcommand{\ee}{\end{eqnarray}}
\newcommand\del{\partial}
\newcommand\nn{\nonumber}
\newcommand{\Tr}{{\rm Tr}}
\newcommand{\mat}{\left ( \begin{array}{cc}}
\newcommand{\emat}{\end{array} \right )}

\title{SURPRISES FOR QCD AT NONZERO CHEMICAL POTENTIAL}

\author{K. Splittorff$^*$ and J.J.M. Verbaarschot$^{**}$}

\address{Niels Bohr Intitute,\\
Blegdamsvej 17, Copenhagen \\
$^*$Email: split@nbi.dk\\
$^{**}$On leave from Stony Brook University, Stony Brook, NY\,11794.\\
Email: jacobus.verbaarschot@stonybrook.edu}

\begin{abstract}

In this lecture we compare different QCD-like  partition functions 
with bosonic quarks and fermionic quarks at nonzero chemical potential. 
Although it is not a surprise that the ground state properties of a 
fermionic quantum system and a bosonic quantum system are completely 
different, the behavior of partition functions with bosonic quarks 
does not follow our naive expectation. Among other surprises, we find 
that the partition function with one bosonic quark only exists at
nonzero chemical potential if a conjugate bosonic quark and a conjugate
fermionic quark are added to the partition function.

\end{abstract}


\bodymatter

\section{Introduction}\label{sec1}

The QCD phase diagram in the chemical potential temperature plane
has far reaching phenomenological implications ranging from 
heavy ion collisions to the interior of neutron stars. Unfortunately,
first principle lattice simulations are only possible at zero 
chemical potential, 
and our knowledge of the phase diagram mainly relies
on model calculations (see for example \refcite{adam,krishna}).   
Physically, we know that at zero temperature the baryon density
is zero below a chemical potential equal to $m_N/3$. Therefore, 
in the thermodynamic limit, the
QCD free energy and its derivatives, such as for example
the chiral condensate, do not depend on the chemical potential for
$\mu <m_N/3$. Since the Dirac operator depends
on the chemical potential this requires miraculous cancellations in
the microscopic theory a problem that was coined\cite{cohen} as the
{\it The Silver Blaze Problem}. This problem becomes particularly
manifest in terms of the eigenvalues of the Dirac operator which are
distributed homogeneously in a strip\cite{all} 
with a width that increases a function of $\mu$.
 \begin{figure}[!t]
         \centerline{
           \scalebox{0.3}{
             \input{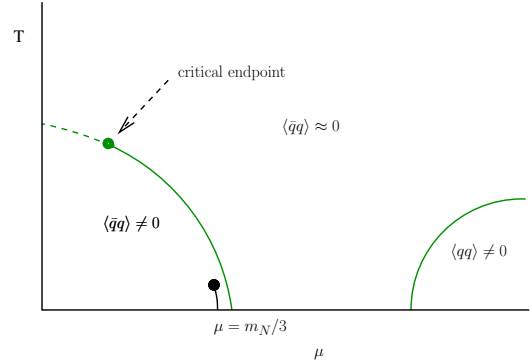} }}
         \caption{Possible phases of the QCD partition function at nonzero
temperature and chemical potential.}
         \label{fig-eg}
       \end{figure}
In this lecture we will mainly focus on the zero temperature axis of the
phase diagram. To better understand the effect of the baryon chemical potential
in QCD, we will consider four different partition
functions listed in Table 1 which were discussed in
 \refcite{OSV} and \refcite{SplitVbos}.
 \begin{table}[h]
\label{table:summary}
\begin{tabular}{c|c|c}
& & \\
{ \textrm{Theory}} & {\textrm{ Number of Charged 
}} 
&{ \textrm{ Critical Chemical}}\\
& {\textrm{  Goldstone Modes }}  &  Potential\\ 
& for $\mu < \mu_c$& \\
 & & \\[-0.3cm]
\hline
&&\\[-0.3cm]
{\large $\langle\det(D+\mu\gamma_0+m)\rangle$} & { \large 0} 
&\bf \large  $\mu_c = \frac 13 m_N $\\
&&\\[-0.3cm]
\hline
&&\\ [-0.3cm]
{\large $\langle|\det(D+\mu\gamma_0+m)|^2\rangle \ \ \ $} & {\large  
$2 $}
 & {\bf \large $\mu_c = \frac 12 m_\pi$} \\
&&\\[-0.3cm]
\hline
&&\\[-0.3cm]
{\large $\langle\frac{1}{\det(D+\mu\gamma_0+m)}\rangle$} & { \large $4$} 
& {\bf \large $\mu_c = \frac 12 m_\pi$} \\
&&\\[-0.3cm]
\hline
&&\\[-0.3cm]
{\large $\langle\frac{1}{|\det(D+\mu\gamma_0+m)|^2}\rangle$} & { \large 
na}
 & {\bf \large $\mu_c = 0$} \\
&&\\
\end{tabular}
{{\small Table 1.
Summary of properties of low energy QCD at nonzero chemical potential and
zero temperature. These partition functions will be denoted by
$Z^{N_f=1}$, $Z_{n=1}$, $Z^{N_f=-1}$, $Z_{n=-1}$, in this order. }}
\end{table}
Although, the first partition function is physically the most relevant,
 the other partition functions have important applications.
Because lattice QCD simulations of full QCD at nonzero chemical potential
are not possible, one sometimes uses the phase quenched approximation
where
\be
\langle {\det}^2(D +\mu \gamma_0 +m) \rangle \to 
\langle |{\det}(D +\mu \gamma_0 +m)|^2 \rangle,
\ee
which can be interpreted as  a partition function of quarks and conjugate 
quarks\cite{misha}.
Then Goldstone bosons made out of quarks and conjugate anti-quarks have
nonzero baryon number resulting in a critical chemical potential of
$m_\pi/2$ instead of $m_N/3$.

The bosonic partition function occurs in the formula for the quenched 
spectral
density in the microscopic domain of QCD which is given by\cite{SplitVerb2}
\be
\rho^{\rm quen}(z,\mu) 
= \frac {|z|^2}2 Z_{n=1}(z,\mu) Z_{n=-1}(z,\mu).
\label{toda}
\ee

In a future publication\cite{splitVphase} 
we will consider the expectation value of the phase of the fermion determinant
given by
\be
\langle e^{2i\theta} \rangle = \left \langle \frac{\det(D +\mu\gamma_0 +m)}
{\det(-D +\mu\gamma_0 +m^*)} \right \rangle.
\label{phase}
\ee
This partition function is not among the above list, but based on our
insights from the bosonic partition function, we will be able to predict
the its phase diagram.

\section{Gauge Invariance and the Phases of QCD at $\mu \ne 0$}\label{sec2}

The principle that underlies the independence of the free energy
on the chemical potential is gauge invariance \cite{GL,KST}. 
The Dirac operator can be
written as
\be
D+\mu\gamma_0 + m = e^{-\mu\tau} (D+m) e^{\mu\tau},
\label{gauge}
\ee
which implies that the $\mu$-dependence can be transformed into
the boundary conditions. A $\mu$-independent free energy is
possible in a phase that is not sensitive to the boundary conditions.
This is the case for $\mu <\mu_c$ when the quarks do not loop around the
torus in the time direction.

Although the partition function $Z_{n=-1}$  is naively gauge invariant
it turns out that the regulator of the partition function breaks 
gauge invariance so that a $\mu$-independent phase cannot exist.
The need for regularization is best seen by writing the partition
function in terms of eigenvalues\cite{AOSV}
\be
Z_{n=-1} = \int_{{\mathbb C}/C_m(\epsilon)} \prod_k d^2z_k \frac{\rho(\{z_k\})}
{\prod_k(z_k^2-m^2)({z^*}_k^2-m^2)},
\ee
where ${\mathbb C}/C_m(\epsilon) $ is the complex plane except two small  
spheres  
with radius $\epsilon$ around $\pm m$. Because of the complex conjugated
pole the integral diverges as $\log \epsilon$. Instead of this 
regularization we prefer to regularize the partition function 
as\cite{SplitVerb2} (also known as 
hermitization\cite{herm})
\be
Z_{n=-1} = \left \langle 
{\det}^{-1} \mat \epsilon & D+\mu\gamma_0+m \\
-D + \mu \gamma_0+m^* & \epsilon \emat \right\rangle.
\label{hermitization}
\ee
Since the matrix inside the determinant is Hermitian, this partition
function can be written as a convergent bosonic integral. However,
$\epsilon$ breaks gauge invariance, and for $\epsilon \ne 0$ it is
not possible to gauge away the chemical potential. We thus find that
$\mu_c = 0 $ in this case.

Let us now consider the partition function with one bosonic flavor.
This partition function {\it cannot}
 be written as a convergent bosonic integral and therefore {\it cannot}
be interpreted in terms of bosonic quarks only.
The correct interpretation is to rewrite this partition function 
as\cite{SplitVbos}
\be
Z^{N_f=-1} = \left \langle \frac {{\det}^*(D+\mu\gamma_0 +m)}
{{\det}(D+\mu\gamma_0 +m){\det}^*(D+\mu\gamma_0 +m)} \right \rangle.
\label{nfm}
\ee
and regulate the denominator as in 
(\ref{hermitization}).
However, contrary to the case of a pair of conjugate bosonic quarks, this
partition function does not diverge for $\epsilon \to 0$, and it is
possible to gauge away the chemical potential. In this case the 
free energy will be $\mu$-independent in the thermodynamic limit 
below the lightest particle
with nonzero baryon number which is a Goldstone boson made out of
a bosonic quark and a conjugate bosonic anti-quark.

\section{Low Energy Limit of QCD}\label{sec3}

The low-energy limit of the partition functions in Table 1 uniquely 
follows from chiral symmetry and gauge invariance. In Table 2 and Fig. 2 
we compare
the bosonic and fermionic (see \refcite{misha,KST,TV,eff,SplitVerb2})
partition functions with a pair of conjugate
flavors. 
 \begin{table*}[h]
{\label{table:summary}}
\begin{tabular}{c|c|c}
& & \\
 & $ \langle |\det(D+\mu\gamma_0+m)|^2\rangle   $ 
& $ \langle |\det(D+\mu\gamma_0+m)|^{-2}\rangle   $\\
 & & \\[-0.3cm]
\hline
&&\\[-0.3cm]
 &no regularization & regularization \\
&&\\ [-0.3cm]
\hline
&&\\ [-0.3cm]
Goldstone          & $U \in U(2)$ & $Q \in Gl(2)/U(2) $ \\
Manifold\footnote{We ignore topological fluctuations.}&&\\&&\\[-0.3cm]
\hline
&&\\[-0.3cm]
Chiral            &$ {\cal L}^{\rm kin} = \frac {F^2}4 {\rm Tr} \nabla_\mu U 
\nabla_\mu U^\dagger$ &${\cal L}^{\rm kin} = \frac{ F^2}4 {\rm Tr} \nabla_\mu Q 
\nabla_\mu Q^{-1} $\\
Lagrangian && \\ && \\[-0.3cm]
\hline
&&\\[-0.3cm]
Covariant  & $ \nabla_0 U = \del_0 U + \mu[U,B]$
&$ \nabla_0 Q = \del_0 Q + \mu\{Q,B\}$\\
Derivative&&\\&&\\[-0.3cm]
\hline
&&\\[-0.3cm]
$\mu_c$ & $\mu_c = \frac {m_\pi}2$& $\mu_c = 0 $\\[0.3cm]
\end{tabular}
\centerline{{\small Table 2.
Comparison of the $n=1$ and the $n=-1$ partition function.\hspace*{2cm}}}
\end{table*}
 \begin{figure}[h]
         \centerline{
           \scalebox{0.45}{
             \input{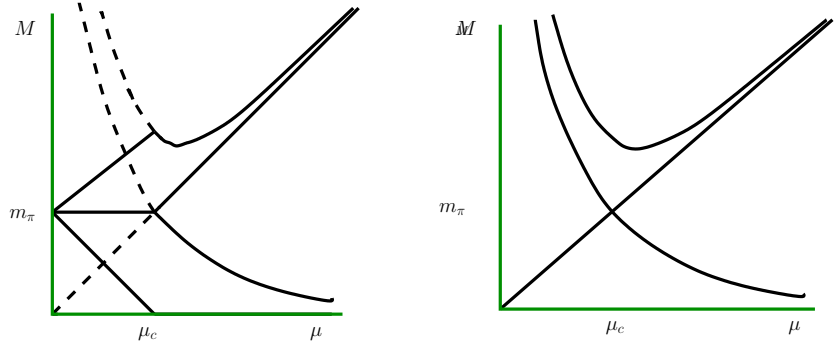} }}
         \caption{Goldstone spectrum for $Z_{n=1}$ (left) and 
$Z_{n=-1}$ (right)}
         \label{fig-eg4}
       \end{figure}

For $N_f =-1$ the partition function is given by the ratio in eq. (\ref{nfm}).
In this case the partition function is finite for vanishing
regulator and the gauge symmetry (\ref{gauge}) is not obstructed. 
Therefore, we have a $\mu$-independent phase for $\mu < \mu_c$. In this
phase the chiral condensate is given by
\be
\langle \bar \psi \psi \rangle^{N_f=-1} &=& \frac 1V\left( 
 \Sigma_k \frac 1{z_k +m}
+\Sigma_k \frac 1{z_k +m} - \Sigma_k \frac 1{z_k +m}\right )
\nn\quad {\rm for} \quad \mu<\mu_c\\
&=&2\Sigma - \Sigma = \Sigma.
\ee
For $\mu>\mu_c$ the bosonic contribution to the chiral condensate rotates
into a pion condensate (see Fig. 3) 
so that for $\mu \gg \mu_c$ only the fermionic contribution remains: 
\be
\langle \bar \psi \psi \rangle = - \Sigma \quad {\rm for} \quad \mu \gg \mu_c.
\ee
 \begin{figure}[h]
         \centerline{
           \scalebox{0.4}{
             \input{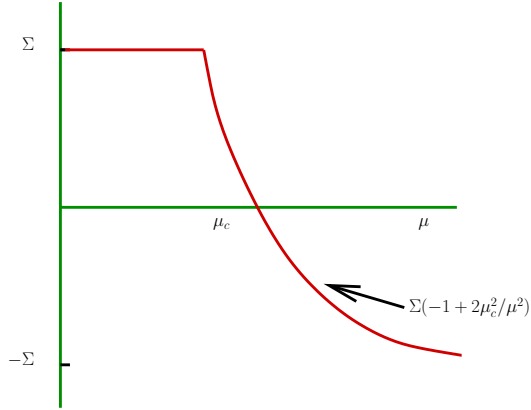} }}
         \caption{The $\mu$-dependence of the chiral condensate for
$N_f=-1$}
         \label{fig-eg5}
       \end{figure}
\section{Chiral Symmetry Breaking at $\mu \ne 0$ and the Dirac Spectrum}
\label{sec4}

In the domain where the kinetic term of the chiral Lagrangian factorizes
from the partition function, i.e. for
$
\mu \ll \     1/L$ and $        |z| \ll  1/{L^2}$,
the quenched spectral density  satisfies the
relation (\ref{toda}). This offers the possibility to test our results
for $Z_{n=-1}$ by means of lattice QCD simulations. Calculations
both with staggered fermions and overlap fermions show an 
impressive agreement\cite{tilo} with (\ref{toda}).

In the thermodynamic limit
the eigenvalues are distributed homogeneously inside the strip
\be
|{\rm Re}(z)| < \frac{2\mu^2 F^2} \Sigma.
\label{strip}
\ee
Therefore, inside this strip, the quenched chiral condensate goes to zero
linearly. 
The behavior of the chiral condensate for full QCD is quite different. In
that case the chiral condensate remains nonzero for $m \to 0$. On the
other hand, the eigenvalues still spread out in the complex plane.
To explain\cite{OSV} 
this so called ``Silver Blaze Problem''
 we introduce the  ``spectral density''
\be
\rho^{\rm full}(z,\mu) = 
\frac{ \langle \det(D+\mu\gamma_0 +m)\sum_k \delta^2(z-z_k) \rangle }
{\langle \det(D+\mu\gamma_0 +m)\rangle }.
\ee
Because of the phase of the fermion determinant $\rho^{\rm full}(z,\mu)$  is
in general complex. 
Its microscopic limit is known analytically\cite{james}  and 
can be decomposed as
\be 
\rho^{\rm full}(z,\mu) = \rho^{\rm quen}(z,\mu)+ \rho^{\rm osc}(z,\mu),
\ee
where $\rho^{\rm osc}$ is complex with
oscillations with a period of $O(1/V)$ and
an amplitude that diverges exponentially with the volume. For $V\to \infty$ 
it vanishes
outside a region  with $ m<    |{\rm Re}(z)|< 
\frac 83 \mu^2F^2/\Sigma - \frac m3$. 
The chiral 
condensate follows the same decomposition
\be
\Sigma^{\rm full} = \Sigma^{\rm quen } + \Sigma^{\rm osc}.
\ee
In Fig. 4 we show the behavior of the different
contributions in the thermodynamic limit. This shows that a nonzero
chiral condensate for $m\to 0$ is due to the oscillatory contribution
to the spectral density\cite{OSV}. Therefore these  oscillations
solve the Silver Blaze Problem. 
 \begin{figure}[!t]
         \centerline{
           \scalebox{0.4}{
             \input{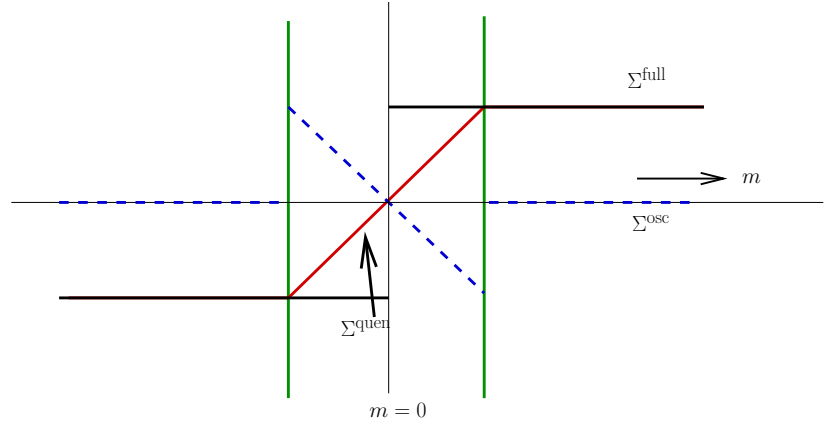} }}
         \caption{The chiral condensate in quenched QCD ($\Sigma^{\rm quen}$)
and in full QCD ($\Sigma^{\rm full}$) as a function of the mass. 
The support of the Dirac spectrum is in between the vertical lines.}
         \label{fig-eg6}
       \end{figure}

\begin{figure}[!h]
         \centerline{
           \scalebox{0.40}{
             \input{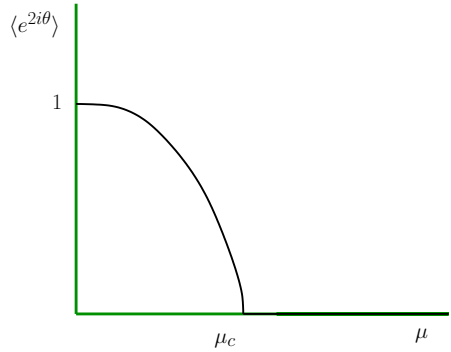} }}
         \caption{Average phase of the fermion determinant
as a function of $\mu$.}
         \label{fig-eg6}
        \end{figure}
\section{Conclusions}\label{sec5}
The behavior of bosonic partition functions at nonzero chemical 
potential is quite different from what could be expected naively.
This surprising behavior can be understood from  gauging
the chemical potential into the boundary 
conditions. 
In particular, this shows that the partition function with a 
pair of conjugate bosonic quarks has no $\mu$-independent phase.
The free energy of the theory with one bosonic quark, on the other hand
is $\mu$-independent for $\mu < m_\pi/2$. However, this partition
function only exists as a partition function of a pair of conjugate bosonic
quarks and a fermionic quark with the same mass. Finally, an analysis along
the lines of this paper\cite{splitVphase} shows that the
expectation value of the phase of the fermion determinant, eq. (\ref{phase}),
behaves as in Fig. \ref{fig-eg6}.

{\bf Acknowledgments}. This work was
supported by DOE Grant No. DE-FG-88ER40388. KS was supported by
the Carslberg Foundation.

\end{document}